\documentclass[journal=jacsat,manuscript=article,layout=twocolumn]{achemso}

\setkeys{acs}{articletitle=true}

\makeatletter
\let\l@addto@macro\relax
\makeatother
\usepackage[fontsize=10pt]{scrextend}

\usepackage[utf8]{inputenc}

\usepackage{graphicx}
\usepackage{subfigure} 
\usepackage{dcolumn}
\usepackage{amsmath}
\usepackage{nameref}
\usepackage{natbib}
\usepackage{bm}
\usepackage{color,soul} 
\usepackage{upgreek}
\usepackage[percent]{overpic}
\usepackage[dvipsnames]{xcolor}
\usepackage[normalem]{ulem}

\usepackage{lipsum}
\usepackage{cuted}
\graphicspath{{figs/}}

\iftrue
  \def\xk#1{\wrapsign{#1}{BlueGreen}{XK}}
  \def\jq#1{\wrapsign{#1}{blue}{JQ}}
  \def\cmt#1{{\color{gray}#1}}
  
  \def\del#1{{\color{red}\sout{#1}}} 
  \def\insert#1{{\color{blue} #1}}
  \def\wrap#1{[\kern-2pt[\thinspace{#1}\thinspace]\kern-2pt]}
  \def\wrapsign#1#2#3{\wrap{{\textcolor{#2}{#1}}}\lower 4pt\hbox{\tiny #3}}
\else
  \def\xk#1{}
  \def\jq#1{}
  \def\cmt#1{}
  
  \def\del#1{}
  \def\insert#1{#1}
\fi

\def\er{\epsilon_{\rm r}}

\def\cF{{\cal F}}

\def\dd{{\rm d}}

\def\ii{{\rm i}}
\def\rv{{{\vv r}}}
\def\qv{{{\vv q}}}
\def\vv#1{\bf{#1}}
\def\sub#1{{\rm #1}}
\title{Microphase separation in neutral homopolymer blends induced by salt-doping}
\author{Xian Kong}
\email{xk@scut.edu.cn}
\affiliation{South China Advanced Institute for Soft Matter Science and Technology, School of Emergent Soft Matter, South China University of Technology, Guangzhou 510640, China}
\affiliation{Guangdong Provincial Key Laboratory of Functional and Intelligent Hybrid Materials and Devices, South China University of Technology, Guangzhou 510640, China}
\author{Jian Qin}
\email{jianq@stanford.edu}
\affiliation{Department of Chemical Engineering, Stanford University, Stanford, CA 94305}

\begin{document}

\maketitle
\begin{abstract}
Microphase separation in polymeric systems provides a bottom-up strategy
to fabricate nanostructures.
\cite{Kim2003Epitaxial,Tavakkoli2012Templating,Suh2017Sub,liu2018directed}
Polymers that are reported to
undergo microphase separation 
usually include block copolymers or polyelectrolytes.
Neutral homopolymers,
which are comparatively easy to synthesize,
are thought to be
incapable of microphase separation.
Here, using a minimal model that
accounts for ion solvation,
we show that
microphase separation is possible in neutral homopolymer blends
with sufficient dielectric contrast,
upon a tiny amount of salt-doping.
The driving force for the microphase separation is
the competition between
selective ion solvation, that
places smaller ions in domains with higher dielectric constant,
and the propensity for local charge neutrality
to decrease the electrostatic energy.
The compromise is an emergent length over which
microphase separation occurs and ions are selectively solvated.
The factors affecting such competitions are explored,
including ion solvation radii, dielectric contrast,
and polymer fraction,
which point to directions for observing
this behavior experimentally.
These findings suggest
a low-cost and facile alternative to produce microphase separation 
which may be exploited in advanced material design and preparation.
\end{abstract}

\section{Background}
Microphase separation results from
the competition among multiple length-dependent 
interactions.\cite{Shi2021Frustration} 
A well-known example is block polymer,
which is formed by chemically linking
thermodynamically incompatible components.\cite{Bates2017Anniversary}
The incompatible blocks tend to separate from each other
while the chemical linkages between blocks prevent macroscopic separation,
giving rise to microphase separation.\cite{Leibler1980Theory}
In the past,
polymeric systems consisting of charged homopolymers
(or polyelectrolytes, PE) 
are also shown to be capable of microphase separation.
One example is weakly charged PEs in poor solvent.
\cite{Dormidontova1994MicrophaseSI,Borue1988statistical,
Joanny1990Weakly,Gritsevich2008Phase,Rumyantsev2017Two}
The incompatibility between PE and solvent promotes a phase separation 
into two macroscopic phases with high and low PE concentrations.
However, this leads to the loss in the translational entropy of counter ions, which mostly reside in the concentrated phase.
The competition between the incompatibility-induced demixing
and translational entropy loss of counter ions,
leads finally to the microphase separation.
Recently, it is suggested
that microphase separation is also possible in polyelectrolytes blends,
\cite{Rumyantsev2020Microphase,Rumyantsev2019Electrostatically}
where the macroscopic 
phase separation between immiscible polyelectroytes
is suppressed by the need to minimize coulombic interaction.

The analogy between the microphase separation of polyelectrolytes and diblock copolymer melts highlights
the role of electrostatic interactions.
Indeed, it has been found that
electrostatic interaction can be leveraged to manipulate the phase behavior of
polyelectrolytes solutions,
\cite{Sing2020Recent,Nakamura2010Self,Nakamura2016Spinodal}
ionic polymer blends,
\cite{Sing2013Interfacial,Sing2013Ion,Pryamitsyn2017Anomalous}
neutral diblock copolymer melts,
\cite{sing2014electrostatic,Kong2021Weakening} 
or charged polymer blends
\cite{Grzetic2021Electrostatic,Fredrickson2022Ionic}.
For example, selective solvation of doped salts in dielectrically heterogeneous copolymers 
has been shown to enhance the effective 
Flory-Huggins parameter between two blocks of diblock copolymer,
\cite{wang2008Effects,Nakamura2011Thermodynamics}
favoring the formation of ordered microscopic phases.
Remarkably, a ``chimney" region was predicted where the solvation effects and
electrostatic correlations of ions can promote microphase formation in an otherwise fully compatible diblock copolymer blends, that is, when the two blocks are fully
miscible.\cite{sing2014electrostatic}
Similarly in polyelectrolyte,
counterions of PEs are predicted to be important in 
determining phase behavior, including enhancing the compatibility between two PEs,
\cite{Zhang1990Intermolecular,Zhang1990NMR}
narrowing the parameter space for microphase separation, 
and allowing the competition between microphase and macrophase separation, etc.
\cite{Fredrickson2022Ionic}

In this work, we show that selective ion solvation can be used
not only to tune the microphase separation of neutral block copolymers
or polyelectrolytes, 
but also to {\it induce} such transitions in neutral homopolymer blends.
We develop a mean-field theory for salt-doped neutral homopolymer blends.
The theory includes the Born solvation and electrostatic interactions of doping salt ions 
in addition to a free energy for neat homopolymer blends.
\cite{Hou2018Solvation,Kong2021Weakening}
A key feature of the theory is the heterogeneous dielectric constant, which depends on the local polymer fraction.
With this theory, we find that under favorable conditions,
the selective solvation of doped ions in domains
with high dielectric constant causes
a microscopic phase transition,
in order to reduce
the loss of translational entropy of ions.

\section{Model and Theory}
We consider binary blends of homopolymers
A and B doped with salts.
The degree of polymerization of the two polymers
are $N_{\rm A}$ and $N_{\rm B}$.
For simplicity,
we only consider the case with one type of salt 
containing one cation species ($+$) and one anion species ($-$).
The valencies of the cation and anion
are $z_+$ and $z_-$, respectively.
In a system containing $n_{\rm A}$ chains of polymer A, 
$n_{\rm B}$ chains of polymer B, $n_+$ cations, and $n_-$ anions, we can define the microscopic volume fraction $\hat{\phi}_i$ (and number density $\hat{\rho}_i$) of each component as,

\begin{align}
\hat{\phi}_{\rm A} (\vv r) &= v_{\rm A} \hat{\rho}_{\rm A} ({\vv r})
= v_{\rm A} \sum_{j=1}^{n_{\rm A}} \int_0^{N_{\rm A}} \dd s \, \delta \left (
{\vv r} -  {\vv r}_j^{\rm A}(s)
\right ) ,
\\
\hat{\phi}_{\rm B} (\vv r) &= v_{\rm B} \hat{\rho}_{\rm B} ({\vv r})
= v_{\rm B} \sum_{j=1}^{n_{\rm B}} \int_0^{N_{\rm B}} \dd s \, \delta \left (
{\vv r} - {\vv r}_j^{\rm B}(s)
\right ) ,
\\
\hat{\phi}_{\rm +} (\vv r) &= v_{\rm +} \hat{\rho}_{\rm +} ({\vv r})
= v_+ \sum_{j=1}^{n_+} \delta \left (
{\vv r} - {\vv r}_j^+
\right ) ,
\\
\hat{\phi}_{\rm -} (\vv r) &= v_{\rm -} \hat{\rho}_{\rm -} ({\vv r})
= v_- \sum_{j=1}^{n_-} \delta \left (
{\vv r} - {\vv r}_j^-
\right ) .
\end{align}
Here,
$v_\alpha$ with $\alpha \in \{ {\rm A}, {\rm B}, +, -\}$ 
are the reference volume\insert{s}
for each component.
The contour curves ${\vv r}_j^p(s)$, 
with $p \in \{ {\rm A}, {\rm B}\}$,
represent the conformation
of chain $j$, in which
$s$ is the contour variable for monomers. 
By writing this, we treat the polymer as a continuous Gaussian chain.
The positions of cation and anion are denoted as
${\vv r}_j^+$ and ${\vv r}_j^-$, respectively.
The microscopic density is evaluated
using the Dirac delta function
$\delta (\vv r)$,
which is normalized and vanishes unless the argument equals~$0$.
In addition,
the salt-doping level is conventionally quantified by
the ratio between 
the number of cations and that of monomers of polymer with high dielectric constant,
$r=n_+/(n_{\rm A}N_{\rm A})$.

To describe interaction in the blends, 
we use a minimal Hamiltonian that is capable 
of reproducing experimental phase diagrams 
of salt-doped diblock copolymer.
\cite{Hou2018Solvation,Hou2020Comparing,Kong2021Weakening} 
The Hamiltonian is composed of four terms 
and written explicitly as,
\begin{align}
\label{eq:hamil}
\beta \cal H =&
\beta({\cal H}^{\rm id}+U^{\rm FH}+U^{\rm B}+U^{\rm C})
\nonumber \\
=&
\sum_{j=1}^{n_{\rm A}} \int_0^{N_{\rm A}}  \frac{3}{2b_{\rm A}^2}
\left( \frac{\dd \rv_j^{\rm A}(s)}{\dd s} \right)^2 \dd s
\nonumber \\
&+\sum_{j=1}^{n_{\rm B}} \int_0^{N_{\rm B}}  \frac{3}{2b_{\rm B}^2}
\left( \frac{\dd \rv_j^{\rm B}(s)}{\dd s} \right)^2 \dd s
\nonumber \\
&+\frac{1}{v_0}\int \dd \rv \chi_\sub{AB}\hat \phi_\sub A(\rv) \hat \phi_\sub B(\rv)
\nonumber \\
&+\int \dd \rv \frac{l_0}{2 \hat \epsilon_{\rm r}(\vv r)}
\left(
\frac{\hat \phi_+(\rv)}{ v_+ a_+} +
\frac{\hat \phi_-(\rv)}{ v_- a_-}
\right)
\nonumber \\
&+
\frac{1}{2}
\iint \dd \rv \, \dd \rv' \,
\hat \rho_{\rm q}(\rv) g(\rv,\rv') \hat \rho_{\rm q}(\rv')
.
\end{align}
Here, $\cal H^{\rm id}$ is the
Hamiltonian of ideal Gaussian chains
that accounts for the conformational statistics
\cite{FredricksonBook}.
$U^{\rm FH}$ is the Flory-Huggins interaction 
between two types of monomers,
and $\chi_{\rm AB}$ is defined on a per-reference volume ($v_0$) basis.

The third term is for ion solvation.
The interaction between ions and polymers is included in solvation free energy
$U^{\rm B}$
approximated using the Born solvation model,\cite{Wang2010Fluctuation}
and no
dispersion interaction among ions is considered.
The terms $l_0 \equiv e^2/4\pi \epsilon_0 k_{\rm B} T$,
$v_i$, and $a_i$
are the vacuum Bjerrum length,
ion volume, and ion diameter, respectively.
The dielectric constant $\hat \epsilon_{\rm r}(\vv r)$
is inhomogeneous and depends on the polymer composition at $\vv r$.
In this work, we use a volumetric mixing rule 
for local dielectric constant,
$\hat \epsilon_{\rm r}(\vv r)=
(\hat \phi_\sub A(\rv) \epsilon_{\rm r, A} + 
\hat \phi_\sub B(\rv)) \epsilon_{\rm r, B}
)/(\hat \phi_\sub A(\rv)+\hat \phi_\sub B(\rv))$,
where $\epsilon_{\rm r, A}$ and $\epsilon_{\rm r, B}$ are
dielectric constant in pure homopolymer A and B.
The ionic contributions to the dielectric constant are neglected,
as we only consider situations with dilute salt contents.

The last term $U^{\rm C}$ is
the Coulombic interaction of the net charge distribution,
$\hat \rho_{\rm q} ({\vv r}) \equiv 
z_+ \hat \rho_+({\vv r}) + 
z_- \hat \rho_-({\vv r})
=
z_+ \hat \phi_+({\vv r})/v_+ + 
z_- \hat \phi_-({\vv r})/v_-
$, in which $g(\rv,\rv')$
is the Green's function for Poisson's equation 
with an inhomogeneous dielectric constant profile,
\begin{align}
-\frac{1}{4\pi l_0} \nabla \cdot \er ({\vv r})
\nabla g(\rv,\rv') = \delta^3(\rv - \rv') .
\end{align}

A key feature of our model related to Born and Coulomb terms is that 
the dielectric constant is inhomogeneous
and depends on
local polymer compositions $\phi_\alpha({\vv r})$,
$\epsilon_{\rm r} ({\vv r})=f[\phi_{\rm A}(\vv r)
, \phi_{\rm B}(\vv r))]$.
As will become clear in the following, 
the selective ion solvation
drives the microphase separation.
It is therefore necessary to point out that 
the Born solvation free energy of an ion of species $\alpha$,
$\sim 1/(a_\alpha \epsilon_{\rm r} ({\vv r}))$,
is inversely proportional to ion radius $a_\alpha$ 
and local dielectric constant $\epsilon_{\rm r} ({\vv r})$.
By this term alone,
we may deduce that ions prefer to stay in domains
with high dielectric constant 
and that the behavior of smaller ion
is more susceptible to the solvation effects.

Following the standard field theory procedures,
\cite{Leibler1980Theory,FredricksonBook,Kong2021Weakening}
which are detailed in Supplementary Information,
we obtain the free energy as functionals of composition fields of all components, $\phi_\alpha(\rv)$ with $\alpha~\in~\{\rm A, B, +, -\}$.
In the disordered phase, the system is uniform and the composition fields are constant, $\phi_\alpha(\rv)=\bar{\phi}_\alpha$.
Therefore the free energy of the disordered phase can be written explicitly and we choose it as a reference state.
For any composition fluctuations around the disordered phase,
we can express the free energy change as Taylor expansion 
in terms of composition differences, 
${\rm \delta} \phi_\alpha(\rv)\equiv \phi_\alpha(\rv)-\bar{\phi}_\alpha$. 
The expansion in Fourier space is written formally as,

\begin{strip}
{
\begin{align}
\label{eq:FEvar4}
\Delta \cF=&\
\Delta \cF^{(2)}+\Delta \cF^{(3)}+\Delta \cF^{(4)}
\nonumber \\=&\
\frac{1}{2V^2} \sum_{\qv} \sum_{\alpha,\beta}
\Gamma^{(2)}_{\alpha\beta}(\qv,-\qv) 
\delta \phi_\alpha(\qv) \delta \phi_\beta(-\qv) \\
& + \frac{1}{3!V^3}\! \sum_{\qv_1,\qv_2} \sum_{\alpha,\beta,\gamma}
\Gamma^{(3)}_{\alpha\beta\gamma} (\qv_1, \qv_2, -\qv_1-\qv_2)
\delta \phi_\alpha(\qv_1) \delta \phi_\beta(\qv_2) \delta \phi_\gamma(-\qv_1-\qv_2) 
  + \cdots, \nonumber
\end{align}
}
\end{strip}
\noindent
where $\alpha,\beta,\gamma,...$ denote species
and $\qv$ is wavevector.
In our convention of Fourier transform,
$
f(\qv)=\int\dd \rv f(\rv) \exp(-\ii \qv \cdot \rv) .
$
The vertex function,
$\Gamma^{(n)}
$, 
is $n$-th order functional derivatives of free energy change 
with respect to density deviations,
evaluated at $\phi(\rv)=\bar \phi$,
or equivalently $\phi(\qv)=\bar \phi \delta (\qv)$.

The secondary expansion coefficient $\Gamma^{(2)}$
is a $4{\times}4$ matrix and its
inverse is the structure factor measured in scattering experiments.
It contains information about the nature of phase separation.
Because of the isotropy of the homogeneous phase,
the elements of $\Gamma^{(2)}$ are functions of $q=|\qv|$,
the wavenumber of density fluctuations.
If we assume the blends are incompressible, 
the composition fluctuations should sum to zero,
$\sum_\alpha \delta \phi_\alpha (\qv)=0$.
This means the composition fluctuation vector
$\delta \phi=
[\delta \phi_\sub A,\delta \phi_\sub B, \delta \phi_\sub +, \delta \phi_\sub -]^{\rm T}$
is orthogonal to a compression mode
${\vv \varepsilon}=\frac{1}{2}[1,1,1,1]^{\rm T}$ and only exist in the orthogonal complement of the subspace of ${\vv \varepsilon}$.
We therefore span the incompressible subspace by choosing
an orthonormal basis set consisting of the following modes,
${\vv e} ^{(1)}=\frac{1}{2}[1,1,-1,-1]^{\rm T}$,
${\vv e} ^{(2)}=\frac{1}{2}[1,-1,1,-1]^{\rm T}$,
and
${\vv e} ^{(3)}=\frac{1}{2}[1,-1,-1,1]^{\rm T}$.
After contracting the composition fluctuations 
to the incompressible subspace,
we reduce the $4{\times}4$ matrix of $\Gamma^{(2)}(q)$ 
to a $3{\times}3$ matrix of $\gamma^{(2)}(q)$.

$\gamma^{(2)}(q)$ should have three eigenvalues $\lambda_1 \leq \lambda_2 \leq \lambda_3$.
In the disordered phase, $\lambda_1(q)>0$, meaning that the free energy is concave up with respect to any composition fluctuation.
The stability limit (i.e. the spinodal limit) of the disordered phase is given by the condition
$\lambda_1(q^*)=0$,
where $q^*$ is a critical wavenumber that minimizes $\lambda_1$.
The nature of the phase separation is indicated by
the value of $q^*$.
If $q^*>0$, the phase transition is microscopic 
and the ordered phase has a characteristic domain size of $D=2\pi / q^*$.
If $q^*=0$, the characteristic domain size diverges 
and the phase separation is macroscopic.

\section{Results and Discussions}

\subsection{Macrophase vs microphase separation}

\begin{figure*}[t]
\includegraphics[width=0.7\textwidth]{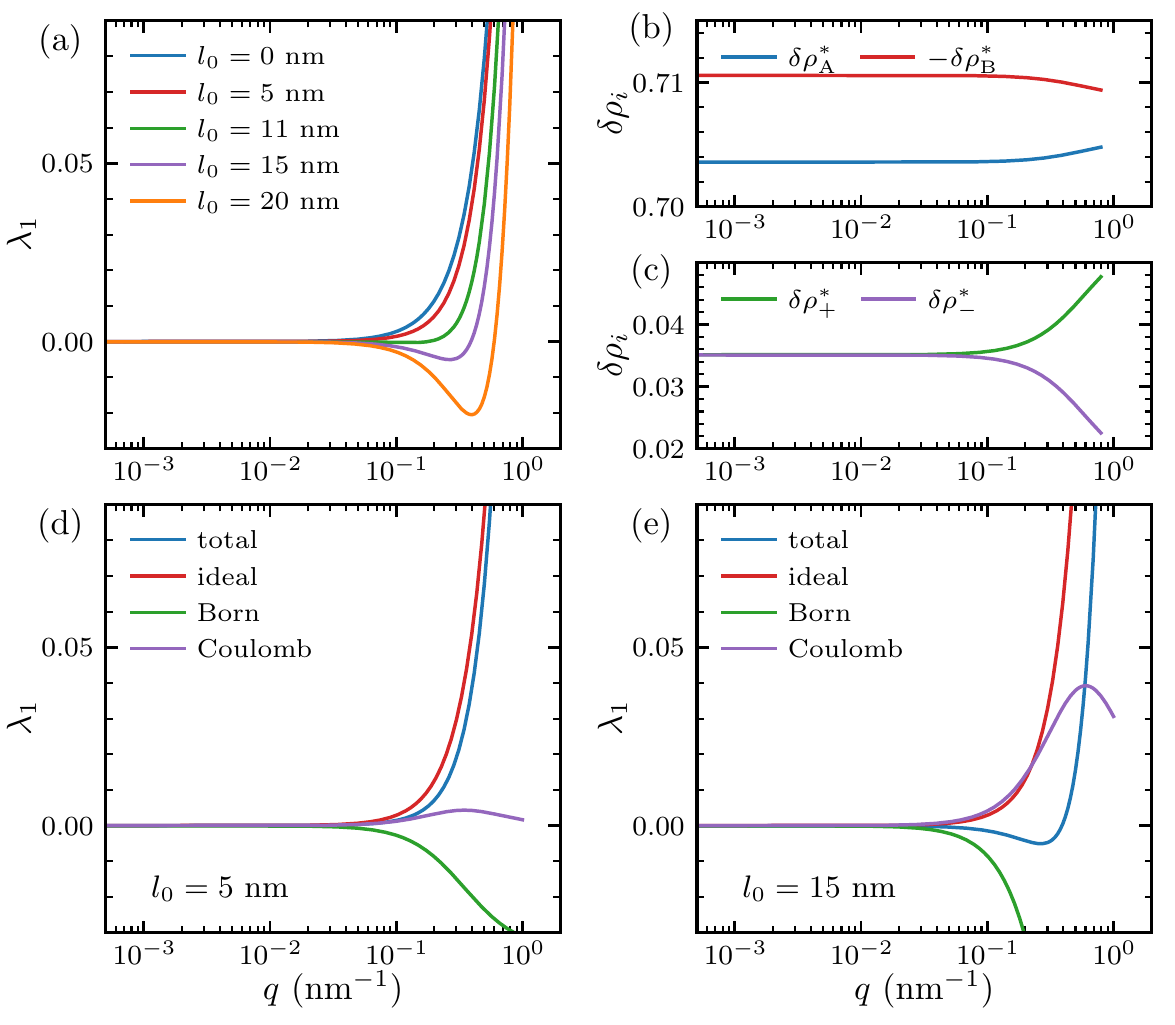}
\caption{
Spectral analysis of the quadratic vertex function of salt-doped homopolymer blends, with
$\chi_{\rm AB}=0$ and $r=0.01$.
(a) Minimum eigenvalue of quadratic coefficients ($\lambda_1$) as function of wavevector magnitude ($q$).
The critical wavevector magnitude ($q^*$) corresponds to the location of the minimum of $\lambda_1$.
The curves were shifted vertically so that the limit values at low $q$ were 0.
(b) Polymeric components of eigenvector corresponding to $\lambda_1$,
with $l_0=15.0\, {\rm nm}$.
(c) Ionic components of eigenvector corresponding to $\lambda_1$,
with $l_0=15.0\, {\rm nm}$.
(d-e) Decomposition of $\lambda_1$ into contributions of ideal entropy, Born solvation, and Coulomb interaction
for $l_0=5.0\, {\rm nm}$ (d) and $l_0=15.0\, {\rm nm}$ (e).
}
\label{fig:spectral}
\end{figure*}

The competition between macroscopic and microscopic phase separation 
is governed by the strength of electrostatic interaction.
We control the electrostatic contributions by tuning the vacuum Bjerrum length $l_0$.
The Bjerrum length at room temperature is about 56 nm in vacuum and 0.7 nm in water.
The (relative)
dielectric constant for polymers is usually small,
between 2 to 10, 
which corresponds to Bjerrum length of about 1.5 to 30 nm.
A small Bjerrum length indicates strong electrostatic screening
and weak electrostatic interactions.

We first consider a symmetric case with $v_{\rm A}=v_{\rm B}=v_0$ and $N_{\rm A}=N_{\rm B}=N$.
Fig.~\ref{fig:spectral}a shows the minimum eigenvalue $\lambda_1$
of $\gamma^{(2)}$ versus $q$ for different Bjerrum lengths.
When $l_0$ is small ($l_0=0 {\text{ or }} 5.0 \,{\rm nm}$),
$\lambda_1$ increases monotonically from $q = 0$.
The minimum of $\lambda_1(q)$ locates at $q^* = 0$.
Therefore only macrophase separation is possible in this regime.
The limiting case of low electrostatic interaction is $l_0=0 \,{\rm nm}$,
where
the electrostatic contribution to the free energy is essentially zero.
This can also be seen from the expressions of the Born solvation energy,
and Coulomb interaction energy (eq.~\ref{eq:hamil}).
In this limit,
the ions act as non-selective neutral solvents and 
it is well-known that only macroscopic separation can occur.

As the value of $l_0$ increases,
$\lambda_1$ starts to change non-monotonically with $q$ (Fig.~\ref{fig:spectral}a).
For $q<0.02\,{\rm nm}^{-1}$, $\lambda_1$
remains flat, 
resembling the
cases with $l_0 \leq 5\,{\rm nm}^{-1}$.
However, beyond $q=0.02\,{\rm nm}^{-1}$, 
the value of $\lambda_1$ first decreases
before finally increasing unboundedly.
This non-monotonic behavior results in a finite critical wavevector $q^*>0$,
signifying a microscopic phase separation.
The value of~$q^*$ increases and
the non-monotonic shape becomes more pronounced
as~$l_0$ increases.

\begin{figure}[tb]
\includegraphics[width=0.45\textwidth]{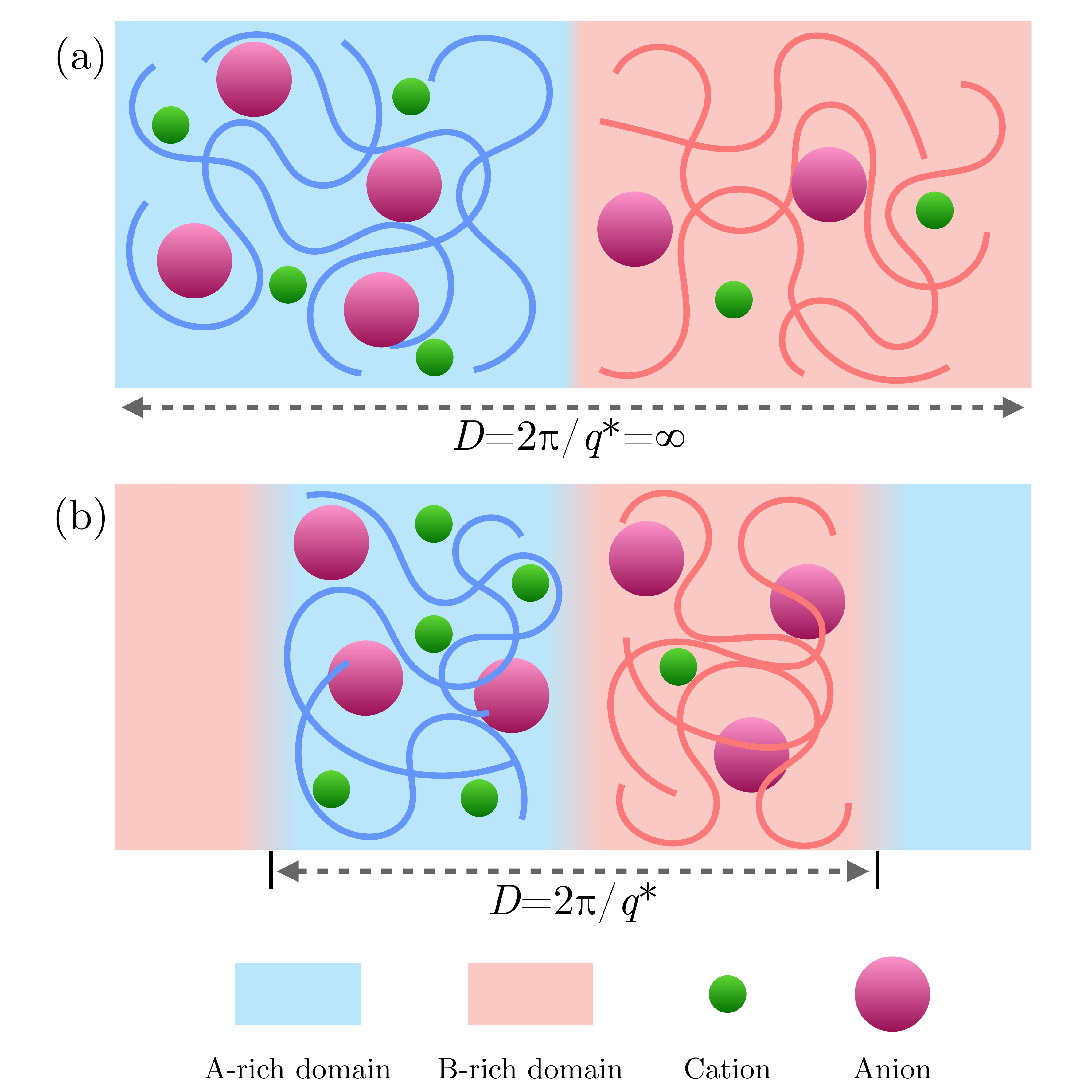}
\caption{
Schematic illustration of the competition between macrophase separation (a) and microphase separation (b).}
\label{fig:schematic}
\end{figure}

To understand the nature of the instability at nonzero~$q^*$,
we examine the components of the critical mode,
i.e., the eigenvector corresponding to the eigenvalue~$\lambda_1$.
The polymeric and ionic components of the critical modes
are plotted in Fig.~\ref{fig:spectral}b and~c, respectively.
To evaluate the volume fractions for different components,
we have used
$\delta \phi^* = \sum_{i=1,2,3} \varphi_i^* {\vv e}^{(i)} $.
The results shown in Fig.~\ref{fig:spectral}b are calculated from $\delta \phi^*$ 
by re-scaling each components with corresponding bead volume, i.e., 
$\delta \rho^*_\alpha=\delta \phi^*_\alpha/v_i$
for $\alpha \in \{{\rm A,B,+,-}\}$.

When $q$ is small ($q<0.02\,{\rm nm}^{-1}$),
$\delta \rho^*_\alpha$ remain approximately constant.
$\delta \rho^*_{\rm A}$ and $\delta \rho^*_{\rm B}$ have different signs
as they tend to separate from each other.
The amplitude $\delta \rho^*_{\rm A}$ is lower than $\delta \rho^*_{\rm B}$,
which is compensated by the enrichment of ions in the~A domain.
The difference between cation and anion number density fluctuation mode 
($\delta \rho _{\rm q}= \delta \rho^*_+ - \delta \rho^*_-$)
measures the degree of net charge separation.
In the regime of $q<0.02\,{\rm nm}^{-1}$,
$\delta \rho^*_+$ and $\delta \rho^*_-$ are essentially identical,
implying the absence of charge separation.
This is consistent with the expectation that charge separation
at large length scale
requires high energy cost.


In the high-$q$ regime,
the magnitudes of
$\delta \rho^*_{\rm A}$ and~$\delta \rho^*_+$ increase,
whereas those of
$\delta \rho^*_{\rm B}$
and $\delta \rho^*_-$ decrease.
This suggests that more cations are distributed in the A-rich domain
while fewer anions reside in the A-rich domain.
It is energetically favorable as 
the small diameter of cation affords a high (absolute value) Born solvation energy
that overcompensates the loss of Born solvation energy 
from the anions that transferred to the B-rich domain.
As $\delta \rho^*_+$ and $\delta \rho^*_-$ split, 
a net charge distribution also develops.
The length scale at which charge separation begins to appear
is about~$\rm 60\,nm$,
calculated from~$2\pi / q^*$ by setting~$q^* = 0.1\,{\rm nm}^{-1}$,
which is well within the range of Coulomb interaction.

To gain more insights into the origin of charge separation,
we decompose $\lambda_1$ into contributions from the ideal, Born and Coulomb parts
(Figs.~\ref{fig:spectral}d, e).
The Flory-Huggins term is irrelevant
because it is $q$-independent and
we set $\chi_{\rm AB}=0$.
All these terms remain constant in the small~$q$ regime.
The ideal part is similar for cases with
$l_0=5\, {\rm nm}$ and $l_0=15\, {\rm nm}$.
(Note that they are not identical, as
their critical composition fluctuations differ slightly.)
In the high-$q$ regime where charge separation takes place (Fig.~\ref{fig:spectral}c),
we find that the decrease in the free energy
is dominated by the
decrease in the Born term (Figs.~\ref{fig:spectral}d,e).
Furthermore, the Coulomb energy increases as the net charge developed.
At even higher $q$ values, the Coulomb contributions decrease.
This is because the
total Coulomb energy decreases as the length scale
of charge separation decreases.

The competition between the Born term and the ideal term 
contributes to the non-monotonic trend in $\lambda_1$
at large $q$ values.
The value of $l_0$ controls the magnitude of Born and Coulomb terms.
Only for sufficiently large~$l_0$ values, can
the Born solvation term dwarfs the ideal term that
causes~$\lambda_1$ to increase, resulting in a well-defined minimum.

The above discussions demonstrate that strong electrostatic interaction can trigger 
the microphase separation in homopolymer blends,
as a result of the competition among multiple factors.
The Born solvation promotes the localization
of ions inside domains with higher dielectric permittivity,
which drives phase separation so that
the high permittivity domains can be formed.
When this happens, both cations and anions
tend to reside inside the high-permittivity phase,
at the cost of the loss
in translational entropy.
One way to alleviate this frustration
is to have smaller cations reside inside
the high-permittivity domain, while allowing
the anions to leak into the low-permittivity domains.
However, this scenario implies macroscopic charge separation,
which is energetically unfavorable:
let the length scale for charge separation be~$D$,
then the magnitude of
net charge~$Q$ is proportional to~$D$,
and the Coulomb energy is of
order~$Q^2 l_0 / D \propto l_0 D$,
which blows up as~$D \to \infty$.
The compromise leads to the emergence of
a finite domain size~$D$, or~$q^*$ value,
when electrostatic interaction is sufficiently strong.
The argument is illustrated in Fig.~\ref{fig:schematic}.
The crossover from macrophase separation
to microphase separation is identified as
the Lifshitz point.
\cite{Rumyantsev2017Two}
The variation of the Lifshitz point and
its dependence on model parameters are explored below.

\subsection{Lifshitz point}

\begin{figure}[tb]
\includegraphics[width=0.4\textwidth]{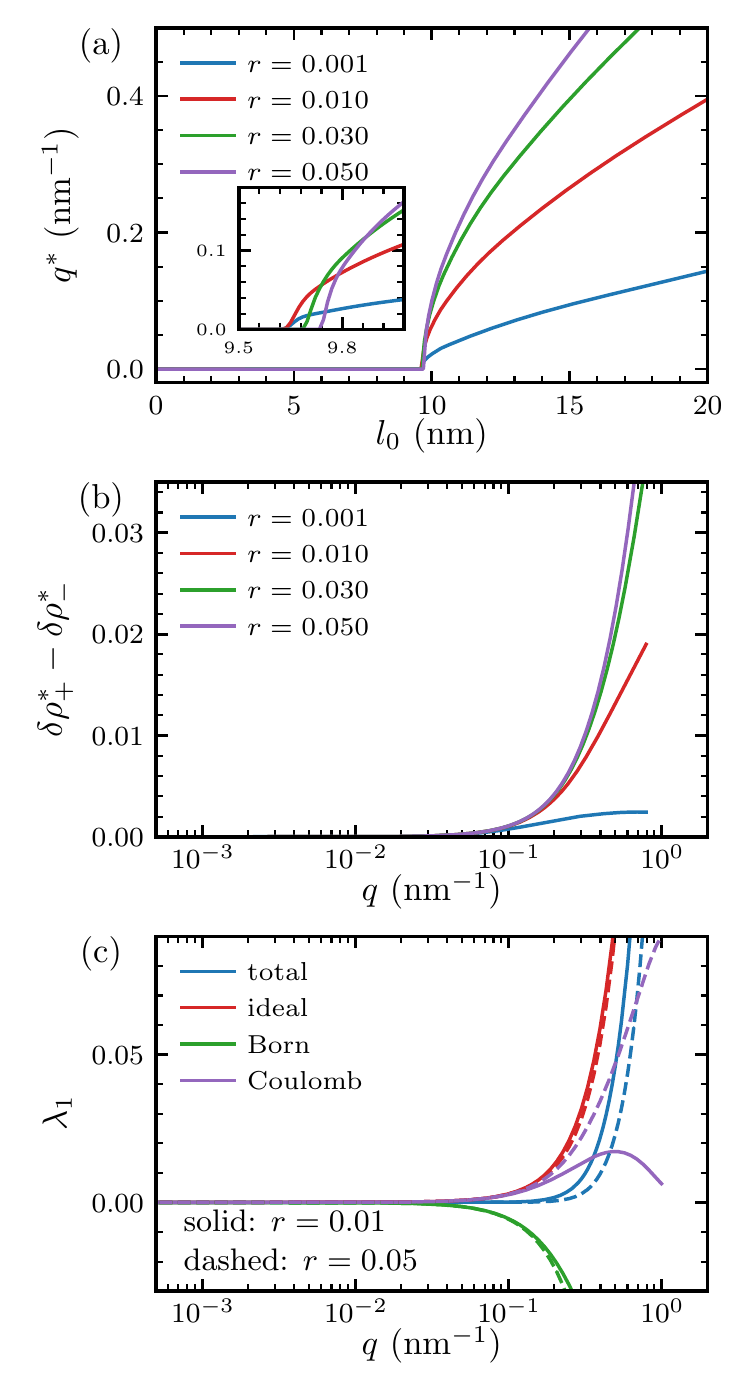}
\caption{
Dependence of critical $l_0$ on salt-doping amounts.
(a) 
Critical wavevector magnitude ($q^*$) versus $l_0$ for different $r$ values.
$q^*$ corresponds to the location of the minimum of $\lambda_1$ in Fig. \ref{fig:spectral}a.
(b) Salt-exchange degree as quantified by $\delta \rho^*_+ - \delta \rho^*_-$ versus wavevector magnitude for different $r$ values.
(c) Decomposition of $\lambda_1$ into contributions of ideal entropy, Born solvation, and Coulomb interaction
for two $r$ values with $l_0=9.8\, {\rm nm}$.
}
\label{fig:doping}
\end{figure}

When the dielectric ratio $\epsilon_{\rm A} / \epsilon_{\rm B}$ is fixed,
the vacuum Bjerrum length $l_0$ is the primary factor determining
the transition from macrophase to microphase separation.
This is demonstrated in Fig.~\ref{fig:doping}a, which
shows how $l_0$ influences $q^*$, 
the wave number where the minimum of $\lambda_1$ locates (Fig.~\ref{fig:spectral}a),
for different salt-doping levels $r$.
The Lifshitz points, where~$q^*$ first
becomes nonzero, are located near~$l_0 = \rm 9.7 nm$.
With the range of salt doping levels explored, from 0.001 to 0.05, 
the location of Lifshitz point barely moves,
as shown by the inset of Fig.~\ref{fig:doping}a.

The weak dependence of the Lifshitz point on the doping level $r$
stems from the insensitivity of the degree of charge separation to $r$.
Fig.~\ref{fig:doping}b compares
the degree of charge separation,
quantified by the difference between the cationic and anionic components 
of the critical mode,
for different doping levels.
The results for different doping levels
are nearly identical when
$q<0.1\, {\rm nm^{-1}}$,
which results in nearly identical
contributions from the Born solvation term.
In fact, for $q<0.1\, {\rm nm^{-1}}$, 
the contribution of the Born term to $\lambda_1$
are almost indistinguishable
for $r=0.01$ and $r=0.05$ (Fig.~\ref{fig:doping}c).
This is precisely
the range at which $q^*$ rises from 0
to finite values,
which rationalizes why the location
of the Lifshitz point is insensitive
to the value of~$r$.

The magnitude of~$q^*$,
i.e., the characteristic domain size,
does depend on the doping level $r$ for $l_0>10\, {\rm nm}$.
The higher the doping level,
the greater the~$q^*$ value,
as seen from Fig.~\ref{fig:doping}a.
Such difference is also related to the progressively
greater difference in the degree of charge separation
for larger~$q^*$ values, shown in
Fig.~\ref{fig:doping}b.

\subsection{Spinodal curves}
The above sections address the conditions
for microphase separation.
Here we examine the stability limit of the homogeneous phase,
by evaluating the spinodal curves, which is found by requiring that
$\lambda_1(q^*)=0$.
Here we recall that
$\lambda_1$ is the minimum eigenvalue of the quadratic expansion coefficient $\gamma^{(2)}$ 
and $q^*$ is the critical wavevector that gives the minimum of $\lambda_1(q)$.
The Flory-Huggins term,
which was ignored in above sections by setting $\chi_{\rm AB}=0$,
contributes $\delta \phi_{\rm A} \delta \phi_{\rm B} \chi_{\rm AB}$ to $\lambda_1$ for all $q$
\cite{Kong2021Weakening},
where $\delta \phi_\alpha$ is the $\alpha$-component of the eigenvector corresponding to $\lambda_1$.
Therefore, changing $\chi_{\rm AB}$ effectively shifts the curves of $\lambda_1(q)$ vertically,
and the spinodal can be readily found
by requiring that
the minimum of $\lambda_1(q)$ vanishes.

\begin{figure}[bt]
\includegraphics[width=0.4\textwidth]{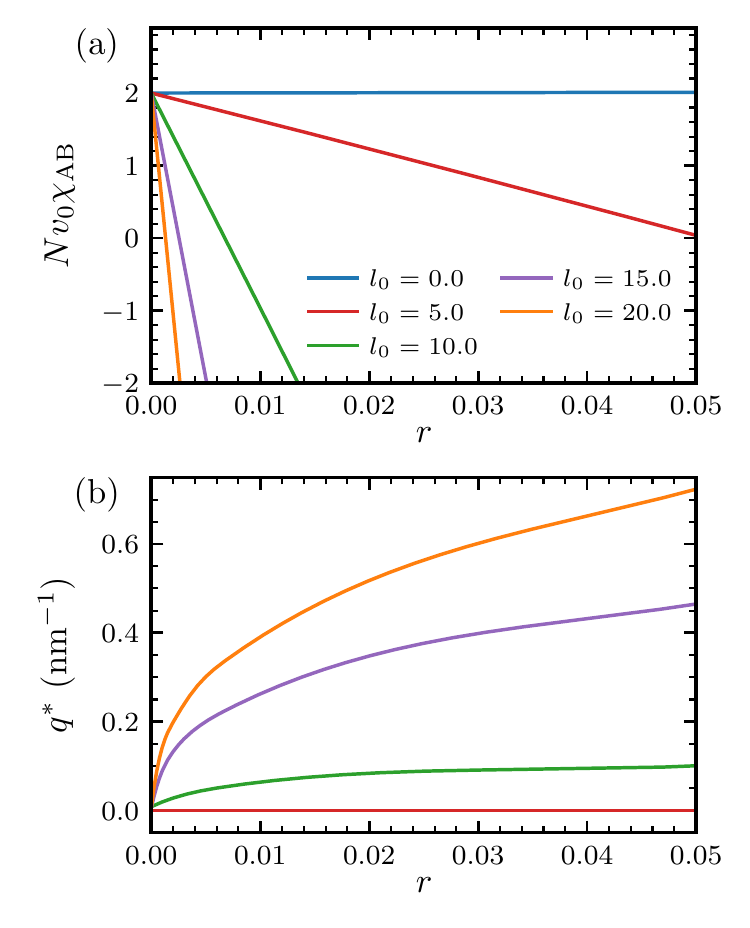}
\caption{
Spinodal behavior of salt-doped homopolymer blends.
(a) 
Spinodal $\chi_{\rm AB}$ values versus doping degree ($r$)
for different $l_0$ values.
(b)
Critical wavevector magnitude ($q^*$) versus $r$ for different $l_0$ values on the spinodal line.
}
\label{fig:spinodal}
\end{figure}

The spinodal values of
$\chi_{\rm AB}$ versus $r$,
for several $l_0$ values,
are plotted in Fig.~\ref{fig:spinodal}a.
All the spinodal curves converge to
$Nv_0\chi_{\rm AB}=2.0$ at $r=0$,
the well-known limit for symmetric binary homopolymer blends
\cite{de1979scaling}.
When $l_0=0\,{\rm nm}$, the ions became essentially non-selective solvents.
The critical composition fluctuation is proportional to $[1,-1,0,0]^{\rm T}$ 
(data not shown),
the same as neat symmetric homopolymer blends.
The value of $\chi_{\rm AB}$
at the spinodal increases slightly with
$r$, because of the dilution effects of non-selective solvents
(the increment is minor for the range of~$r$ is narrow).
For nonzero $l_0$ values,
the value of~$\chi_{\rm AB}$ decreases with~$r$,
and the change
is more substantial for larger $l_0$.
This corroborates the notion
that salt-doping can increase the effective $\chi$ parameter,
as was first proposed by Wang
\cite{wang2008Effects}.
However, Wang mainly
considered the macrophase separation,
whereas our focus is the emergence of microphase separation.

This point is highlighted by the critical wavenumber
at the spinodal shown in
Fig.~\ref{fig:spinodal}b.
Because the Flory-Huggins term does not
alter the~$q$-dependence of~$\gamma_2$,
the information contained in Fig.~\ref{fig:spinodal}c
is the same as that in Fig.~\ref{fig:doping}b.
Taken
together, these results suggest
that the crossover between macro- and micro-phase separation
occurs slightly above~$l_0 = 5\,\rm nm$.

\subsection{Other factors}
Using symmetric homopolymer blends as a model system, 
we have demonstrated the possibility of microphase separation
upon salt-doping and studied the influence of salt content ($r$) 
and electrostatic interaction strength ($l_0$)
in the above.
In the following,
we explore the influences of three key molecular properties: 
ion solvation radius, 
dielectric contrast,
and polymer composition.

\subsubsection{Ion solvation radius}
The driving force for microphase separation
in the system of interests
is the need to simultaneously
lower the Born solvation free energy,
reduce the entropy loss of localized ions,
and minimize the Coulomb energy.
The Born solvation energy,
eq.~(\ref{eq:hamil}),
is inverse to the ion solvation radius.
Accordingly, we expect that the difference in
ion solvation radii of cations and anions to be important.
The parameterization for ion radii
follows our recent study
\cite{Hou2020Comparing}, in which
ionic volumes $v_+$ and $v_-$ are kept constant
and ion radii $a_+$ and $a_-$ are varied.

\begin{figure}[bt]
\includegraphics[width=0.4\textwidth]{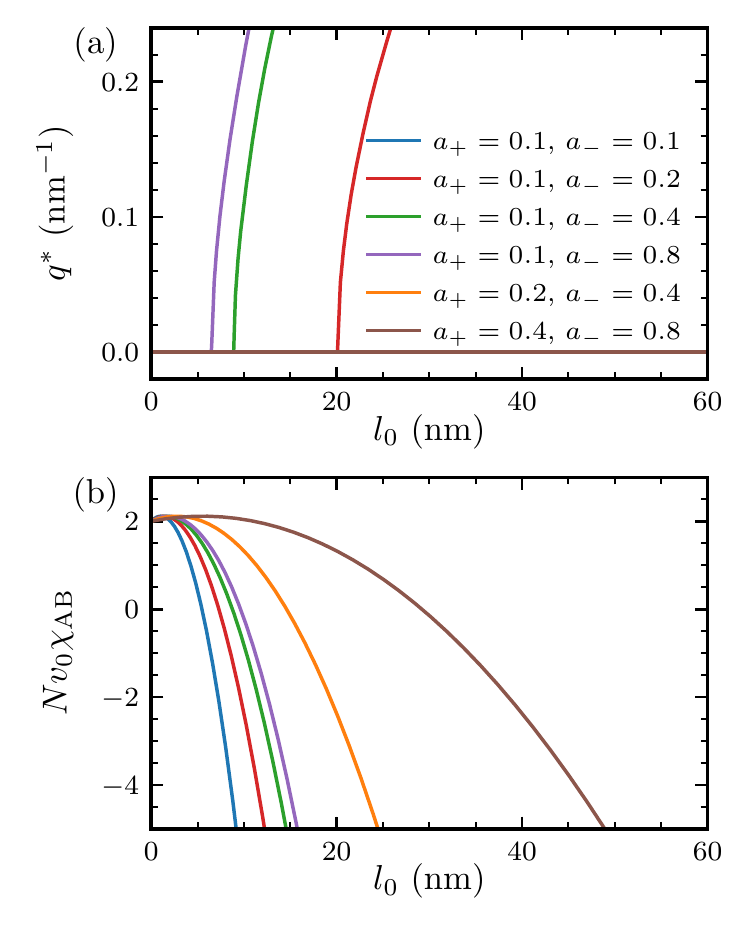}
\caption{
Effects of ionic solvation radii on the phase behavior 
of salt-doped homopolymer blends.
Dependence of critical wavevector magnitude $q^*$ (a)
and spinodal $\chi_{\rm AB}$ on $l_0$ (b)
on $l_0$ for different combinations of ionic solvation radii
($a_+$ and $a_-$).
Doping degree is $r=0.01$.
The unit of ion solvation radius is nm.
}
\label{fig:radius}
\end{figure}

Several combinations of ion solvation radii are explored,
and the critical wavenumbers are plotted
vs~$l_0$ in Fig.~\ref{fig:radius}a.
The symmetric case with $a_+=a_-=0.1\,{\rm nm}$
is indifferent to the selective cation or anion solvation
and charge separation cannot occur,
so there is no
microphase separation for all $l_0$ values.
As $a_-$ increases,
the discrepancy between cation and anion size grows,
and the selective solvation of cations in the high-permittivity domain is stronger.
We found that the value of~$l_0$ at the Lifshitz point decreases
from $\sim$20 nm to $\sim$7 nm
as $a_-$ increases from $\rm 0.2\, nm$ to
$\rm 0.8\, nm$,
which supports our argument that
a large difference between ion radii promotes microphase separation.

Additionally, when the solvation radii are doubled
while the ratio $a_+ / a_-$ is kept constant,
the microphase separation is found to disappear.
Doubling both ion solvation size act effectively
halves the Born term.
This reduction
cannot be offset by simply doubling $l_0$,
as doubling $l_0$ also increases energy cost from the Coulomb term, 
which weakens the energy gained from the Born term,
making microphase separation impossible.

Fig.~\ref{fig:radius}b presents the spinodal curves
for different combinations of ion solvation radii.
The trends of these curves are similar:
the value of~$\chi_{\rm AB}$ decreases
as~$l_0$ increases.
There is, however, a weak increment when $l_0$ is small.
This is the entropy regime of salt-doping,
where adding ions stabilizes the homogeneous phase 
in order to achieve higher translational entropy
\cite{Kong2021Weakening}.
Because no microphase separation is
expected in the entropy regime,
we shall focus on the solvation regime below,
where adding ions de-stabilizes the homogeneous phase.
The decrease of $\chi_{\rm AB}$ with increasing $l_0$ 
depends only on the solvation size of ions 
and is insensitive to the ratio of the ion solvation radii.
Smaller ions decreases $\chi_{\rm AB}$ more effectively,
which is consistent with the findings of Nakamura et al.
\cite{Nakamura2011Thermodynamics}.
We note that such dependence
is irrespective of the nature of phase separation,
macroscopic or microscopic,
consistent with the findings in
Fig.~\ref{fig:spinodal}.

\subsubsection{Dielectric contrast}

\begin{figure}[bt]
\includegraphics[width=0.4\textwidth]{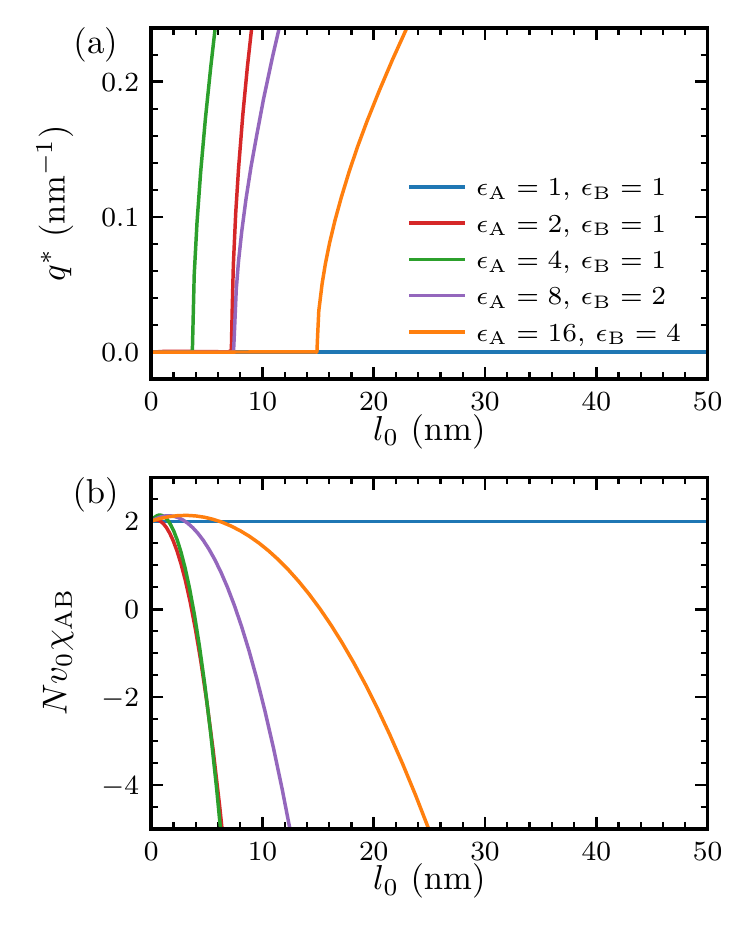}
\caption{
Effects of dielectric constant of polymer on the phase behavior 
of salt-doped homopolymer blends.
Dependence of critical wavevector magnitude $q^*$ (a)
and spinodal $\chi_{\rm AB}$ on $l_0$ (b)
on $l_0$ for different combinations of polymer dielectric constant
($\epsilon_{\rm A}$ and $\epsilon_{\rm B}$).
Doping degree is $r=0.01$.
}
\label{fig:dielectric}
\end{figure}

The effects of dielectric contrast
is analogous to that of the difference in
ion solvation radii.
The values of~$q^*$ and the spinodal curves
are shown in
Fig.~\ref{fig:dielectric}a,
for different sets of dielectric permittivities.
In the absence of dielectric contrast 
($\epsilon_{\rm A}=\epsilon_{\rm B}=1$ in
Fig.~\ref{fig:dielectric}a),
no microscopic separation is found.
Because the net charge density vanishes,
the ions act effectively as neutral solvents.

The microphase separation is possible
with sufficient dielectric contrast.
When we fix the dielectric constant of one polymer ($\epsilon_{\rm B}=1$),
and increase $\epsilon_{\rm A}$ from 2 to 4,
the value of $l_0$ at the Lifshitz point
decreases from $\sim$7.5 nm to $\sim$3.7 nm
(Fig.~\ref{fig:dielectric}a).
However, the spinodal $\chi_{\rm AB}$ does not change significantly from~$\epsilon_{\rm A} = 2$
to~$\epsilon_{\rm B} = 4$
(Fig.~\ref{fig:dielectric}b).
The quantitative dependence should be
sensitive to the average rule chosen
for the dielectric permittivity of mixtures\cite{Kong2021Weakening}.

With a constant $\epsilon_{\rm A}/\epsilon_{\rm B}$,
increasing both dielectric constants shifts
the value of $l_0$ at the Lifshitz point to larger values.
Both Born and Coulomb terms scale inversely
with local dielectric constant.
Doubling both dielectric constants reduces both terms
by a factor of two.
This is equivalent to reducing
the Bjerrum length by a factor of two.
As shown in Fig.~\ref{fig:dielectric}a,
when we change the dielectric constant from 
$(\epsilon_{\rm A},\epsilon_{\rm B})=(4,1)$ to $(8,2)$,
the critical $l_0$ changes
from $\sim$3.7\,nm to $\sim$7.5\,nm.
Further change to $(16,4)$ gives a critical $l_0$ of $\sim$15.0\,nm.

\subsubsection{Polymer composition}

\begin{figure}[bt]
\includegraphics[width=0.4\textwidth]{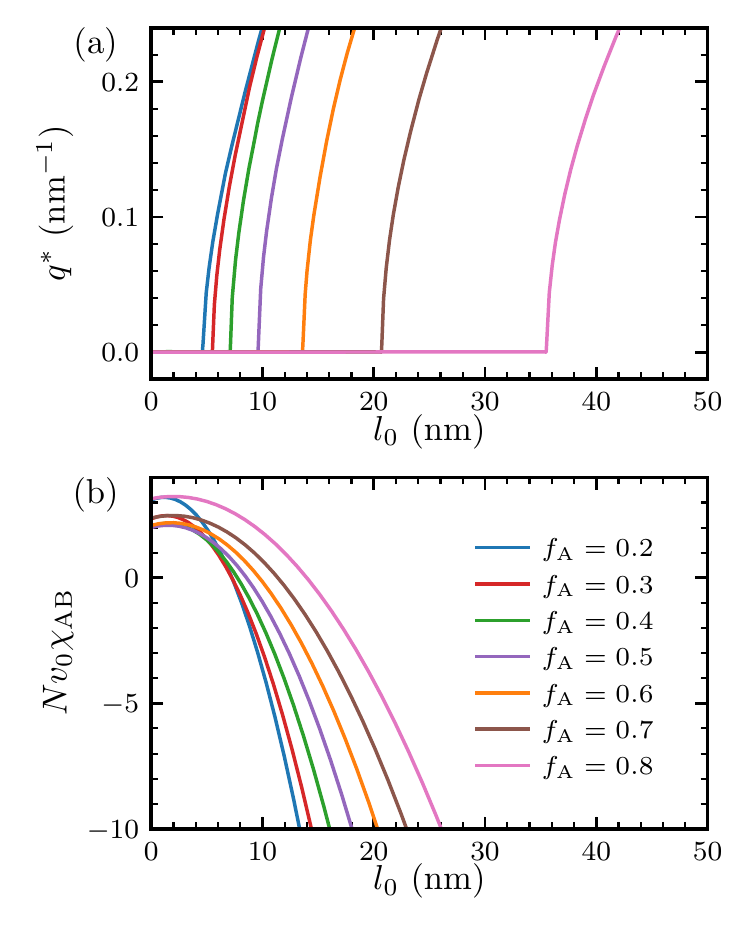}
\caption{
Effects of polymer fraction  on the phase behavior 
of salt-doped homopolymer blends.
Dependence of critical wavevector magnitude $q^*$ (a)
and spinodal $\chi_{\rm AB}$ on $l_0$ (b)
on $l_0$ for different fractions of A-polymer $f_{\rm A}$.
Doping degree is $r=0.01$.
}
\label{fig:fraction}
\end{figure}

Fig.~\ref{fig:fraction}a shows that a higher
fraction of high-permittivity component
($f_{\rm A}$) requires a larger $l_0$ to
attain microphase separation.
This is related
to the translational entropy of ions.
When macroscopic phase separation occurs, ions are distributed more preferably in the A-rich phase.
A small $f_{\rm A}$ means the space available for ions is more restricted and hence a large loss of ionic translational entropy.
The free energy gain is then more likely to drive a transition 
from macrophase separation to microphase separation.

A similar trend is found for how $f_{\rm A}$ affects
the spinodal curves (Fig.~\ref{fig:fraction}b).
When $l_0=0\,{\rm nm}$, spinodal $\chi_{\rm AB}$ values are the same as neat polymer blends,
\cite{de1979scaling}
which is symmetric from $f_{\rm A}=0.5$.
As $l_0$ increases, the value of
$\chi_{\rm AB}$ at the spinodal decreases as expected
but the changing rate is steeper 
for smaller~$f_{\rm A}$ values.
When $f_{\rm A}$ is large,
the average dielectric constant in homogeneous phase
is closer to $\epsilon_{\rm A}$,
and the gain in the Born energy from ion localization is smaller,
weakening the driving force for phase separation
(either macroscopic or  microscopic).

\subsection{Why microphase separation induced by salt-doping seldom seen experimentally?}
The region of microphase separation identified in our theory is rather broad
(Fig.~\ref{fig:spinodal}).
However, such salt-doping-induced microphase separation in neutral homopolymer blends has not been reported.
This may be attributed to
the over-estimation of the solvation free energy
by the simple Born expression.
In reality,
many factors can mitigate ion solvation free energy such as
ion-paring, ion-clustering, and composition fluctuations
\cite{Kong2021Weakening}.
Other treatments of ion solvation 
such as dipolar SCFT,
\cite{Nakamura2014Ion,Nakamura2012Ion,Nakamura2015Dipolar}
liquid state theory corrected SCFT,
\cite{Sing2013Ion,sing2014electrostatic,Sing2013Interfacial}
or classical density functional theory
\cite{Brown2018Ion,Li2006Density,Wu2007Density,Kong2017SLB}
may help improve the accuracy of theoretical predictions 
but will not change our conclusions qualitatively.
Formally weakening the Born solvation term implies that
a higher $l_0$ value is required to induce microphase separation.
To fulfill this condition,
working with low-permittivity polymers
is desirable.
It remains to be seen, if tuning
the ion solvation raii, dielectric contrast,
and blending composition, as we explored above,
can lead to experimental realization of
microphase separation in doped polymer blends.

\section{Conclusions}
We developed a weak segregation theory for
salt-doped neutral homopolymer blends.
The model contains terms that describe 
the solvation free energy of ions and Coulombic interaction
in addition to
standard terms for neutral polymer blends,
which has previously been used to analyze experimental phase diagrams
\cite{Hou2018Solvation,Hou2020Comparing}.
Our main result is that microphase separation
may be induced when selective solvation is sufficiently strong.

The microphase separation permits local charge separation, with cations preferentially
residing in the high-permittivity domains,
whereas anions resides in the low-permittivity domains. 
The net result is that the Born solvation
free energy is lower,
ion entropy loss is reduced,
and the Coulomb energy is minimal.
The threshold value of~$l_0$
at the crossover from the macro- to micro-phase separation is not sensitive to the amount of
salt added.
However, the doping level
changes the critical $q^*$ significantly,
which provides a facile means to tune the domain size of microphases.

We further probed how three key material properties, i.e.,
ion solvation radius, dielectric contrast, polymer fraction
affect our results on microphase separation.
It is found that larger solvation
radius difference,
larger dielectric contrast,
and lower high-permittivity polymer composition,
all favor the formation of microphases.
Such observations may facilitate the
experimental exploration of salt-induced
microphase separation in polymer blends.

This work focused on the competition of microphase and macrophase separation
and the stability limit of the homogeneous phase.
The complete phase diagrams
for salt-doped polymer blends,
including the standard set of microphases
(BCC, hexagonal, gyroid, etc.)
\cite{Leibler1980Theory,Hou2018Solvation, Kong2021Weakening}
will be presented in the future.

\begin{acknowledgement}
This research has been supported by the Recruitment Program of Guangdong (grant no. 2016ZT06C322).
J.Q. is supported by the National Science Foundation 
CAREER Award through DMR-1846547.
\end{acknowledgement}

\bibliography{bib/ref}

\providecommand{\latin}[1]{#1}
\makeatletter
\providecommand{\doi}
  {\begingroup\let\do\@makeother\dospecials
  \catcode`\{=1 \catcode`\}=2 \doi@aux}
\providecommand{\doi@aux}[1]{\endgroup\texttt{#1}}
\makeatother
\providecommand*\mcitethebibliography{\thebibliography}
\csname @ifundefined\endcsname{endmcitethebibliography}
  {\let\endmcitethebibliography\endthebibliography}{}
\begin{mcitethebibliography}{41}
\providecommand*\natexlab[1]{#1}
\providecommand*\mciteSetBstSublistMode[1]{}
\providecommand*\mciteSetBstMaxWidthForm[2]{}
\providecommand*\mciteBstWouldAddEndPuncttrue
  {\def\EndOfBibitem{\unskip.}}
\providecommand*\mciteBstWouldAddEndPunctfalse
  {\let\EndOfBibitem\relax}
\providecommand*\mciteSetBstMidEndSepPunct[3]{}
\providecommand*\mciteSetBstSublistLabelBeginEnd[3]{}
\providecommand*\EndOfBibitem{}
\mciteSetBstSublistMode{f}
\mciteSetBstMaxWidthForm{subitem}{(\alph{mcitesubitemcount})}
\mciteSetBstSublistLabelBeginEnd
  {\mcitemaxwidthsubitemform\space}
  {\relax}
  {\relax}

\bibitem[Ouk~Kim \latin{et~al.}(2003)Ouk~Kim, Solak, Stoykovich, Ferrier,
  De~Pablo, and Nealey]{Kim2003Epitaxial}
Ouk~Kim,~S.; Solak,~H.~H.; Stoykovich,~M.~P.; Ferrier,~N.~J.; De~Pablo,~J.~J.;
  Nealey,~P.~F. Epitaxial self-assembly of block copolymers on lithographically
  defined nanopatterned substrates. \emph{Nature} \textbf{2003}, \emph{424},
  411--414\relax
\mciteBstWouldAddEndPuncttrue
\mciteSetBstMidEndSepPunct{\mcitedefaultmidpunct}
{\mcitedefaultendpunct}{\mcitedefaultseppunct}\relax
\EndOfBibitem
\bibitem[G. \latin{et~al.}(2012)G., Gotrik, Hannon, Alexander-Katz, Ross, and
  Berggren]{Tavakkoli2012Templating}
G.,~A. T.~K.; Gotrik,~K.~W.; Hannon,~A.~F.; Alexander-Katz,~A.; Ross,~C.~A.;
  Berggren,~K.~K. Templating Three-Dimensional Self-Assembled Structures in
  Bilayer Block Copolymer Films. \emph{Science} \textbf{2012}, \emph{336},
  1294--1298\relax
\mciteBstWouldAddEndPuncttrue
\mciteSetBstMidEndSepPunct{\mcitedefaultmidpunct}
{\mcitedefaultendpunct}{\mcitedefaultseppunct}\relax
\EndOfBibitem
\bibitem[Suh \latin{et~al.}(2017)Suh, Kim, Moni, Xiong, Ocola, Zaluzec,
  Gleason, and Nealey]{Suh2017Sub}
Suh,~H.~S.; Kim,~D.~H.; Moni,~P.; Xiong,~S.; Ocola,~L.~E.; Zaluzec,~N.~J.;
  Gleason,~K.~K.; Nealey,~P.~F. Sub-10-nm patterning via directed self-assembly
  of block copolymer films with a vapour-phase deposited topcoat. \emph{Nature
  nanotechnology} \textbf{2017}, \emph{12}, 575--581\relax
\mciteBstWouldAddEndPuncttrue
\mciteSetBstMidEndSepPunct{\mcitedefaultmidpunct}
{\mcitedefaultendpunct}{\mcitedefaultseppunct}\relax
\EndOfBibitem
\bibitem[Liu \latin{et~al.}(2018)Liu, Franke, Mignot, Xie, Yeung, Zhang, Chi,
  Zhang, Farrell, Lai, \latin{et~al.} others]{liu2018directed}
Liu,~C.-C.; Franke,~E.; Mignot,~Y.; Xie,~R.; Yeung,~C.~W.; Zhang,~J.; Chi,~C.;
  Zhang,~C.; Farrell,~R.; Lai,~K., \latin{et~al.}  Directed self-assembly of
  block copolymers for 7 nanometre FinFET technology and beyond. \emph{Nature
  Electronics} \textbf{2018}, \emph{1}, 562--569\relax
\mciteBstWouldAddEndPuncttrue
\mciteSetBstMidEndSepPunct{\mcitedefaultmidpunct}
{\mcitedefaultendpunct}{\mcitedefaultseppunct}\relax
\EndOfBibitem
\bibitem[Shi(2021)]{Shi2021Frustration}
Shi,~A.-C. Frustration in block copolymer assemblies. \emph{Journal of Physics:
  Condensed Matter} \textbf{2021}, \emph{33}, 253001\relax
\mciteBstWouldAddEndPuncttrue
\mciteSetBstMidEndSepPunct{\mcitedefaultmidpunct}
{\mcitedefaultendpunct}{\mcitedefaultseppunct}\relax
\EndOfBibitem
\bibitem[Bates and Bates(2017)Bates, and Bates]{Bates2017Anniversary}
Bates,~C.~M.; Bates,~F.~S. 50th Anniversary Perspective: Block Polymers—Pure
  Potential. \emph{Macromolecules} \textbf{2017}, \emph{50}, 3--22\relax
\mciteBstWouldAddEndPuncttrue
\mciteSetBstMidEndSepPunct{\mcitedefaultmidpunct}
{\mcitedefaultendpunct}{\mcitedefaultseppunct}\relax
\EndOfBibitem
\bibitem[Leibler(1980)]{Leibler1980Theory}
Leibler,~L. Theory of Microphase Separation in Block Copolymers.
  \emph{Macromolecules} \textbf{1980}, \emph{13}, 1602--1617\relax
\mciteBstWouldAddEndPuncttrue
\mciteSetBstMidEndSepPunct{\mcitedefaultmidpunct}
{\mcitedefaultendpunct}{\mcitedefaultseppunct}\relax
\EndOfBibitem
\bibitem[Dormidontova \latin{et~al.}(1994)Dormidontova, Erukhimovich, and
  Khokhlov]{Dormidontova1994MicrophaseSI}
Dormidontova,~E.~E.; Erukhimovich,~I.; Khokhlov,~A.~R. Microphase separation in
  poor‐solvent polyelectrolyte solutions: Phase diagram. \emph{Macromolecular
  Theory and Simulations} \textbf{1994}, \emph{3}, 661--675\relax
\mciteBstWouldAddEndPuncttrue
\mciteSetBstMidEndSepPunct{\mcitedefaultmidpunct}
{\mcitedefaultendpunct}{\mcitedefaultseppunct}\relax
\EndOfBibitem
\bibitem[Borue and Erukhimovich(1988)Borue, and
  Erukhimovich]{Borue1988statistical}
Borue,~V.~Y.; Erukhimovich,~I.~Y. A statistical theory of weakly charged
  polyelectrolytes: fluctuations, equation of state and microphase separation.
  \emph{Macromolecules} \textbf{1988}, \emph{21}, 3240--3249\relax
\mciteBstWouldAddEndPuncttrue
\mciteSetBstMidEndSepPunct{\mcitedefaultmidpunct}
{\mcitedefaultendpunct}{\mcitedefaultseppunct}\relax
\EndOfBibitem
\bibitem[{Joanny, J.F.} and {Leibler, L.}(1990){Joanny, J.F.}, and {Leibler,
  L.}]{Joanny1990Weakly}
{Joanny, J.F.},; {Leibler, L.}, Weakly charged polyelectrolytes in a poor
  solvent. \emph{J. Phys. France} \textbf{1990}, \emph{51}, 545--557\relax
\mciteBstWouldAddEndPuncttrue
\mciteSetBstMidEndSepPunct{\mcitedefaultmidpunct}
{\mcitedefaultendpunct}{\mcitedefaultseppunct}\relax
\EndOfBibitem
\bibitem[Gritsevich(2008)]{Gritsevich2008Phase}
Gritsevich,~A.~V. Phase diagrams of polyelectrolyte solutions in poor solvents
  and of polyelectrolyte globules with allowance for microphase separation and
  fluctuation effects. \emph{Polymer Science Series A} \textbf{2008},
  \emph{50}, 58--67\relax
\mciteBstWouldAddEndPuncttrue
\mciteSetBstMidEndSepPunct{\mcitedefaultmidpunct}
{\mcitedefaultendpunct}{\mcitedefaultseppunct}\relax
\EndOfBibitem
\bibitem[Rumyantsev and Kramarenko(2017)Rumyantsev, and
  Kramarenko]{Rumyantsev2017Two}
Rumyantsev,~A.~M.; Kramarenko,~E.~Y. Two regions of microphase separation in
  ion-containing polymer solutions. \emph{Soft Matter} \textbf{2017},
  \emph{13}, 6831--6844\relax
\mciteBstWouldAddEndPuncttrue
\mciteSetBstMidEndSepPunct{\mcitedefaultmidpunct}
{\mcitedefaultendpunct}{\mcitedefaultseppunct}\relax
\EndOfBibitem
\bibitem[Rumyantsev and de~Pablo(2020)Rumyantsev, and
  de~Pablo]{Rumyantsev2020Microphase}
Rumyantsev,~A.~M.; de~Pablo,~J.~J. Microphase Separation in Polyelectrolyte
  Blends: Weak Segregation Theory and Relation to Nuclear “Pasta”.
  \emph{Macromolecules} \textbf{2020}, \emph{53}, 1281--1292\relax
\mciteBstWouldAddEndPuncttrue
\mciteSetBstMidEndSepPunct{\mcitedefaultmidpunct}
{\mcitedefaultendpunct}{\mcitedefaultseppunct}\relax
\EndOfBibitem
\bibitem[Rumyantsev \latin{et~al.}(2019)Rumyantsev, Gavrilov, and
  Kramarenko]{Rumyantsev2019Electrostatically}
Rumyantsev,~A.~M.; Gavrilov,~A.~A.; Kramarenko,~E.~Y. Electrostatically
  Stabilized Microphase Separation in Blends of Oppositely Charged
  Polyelectrolytes. \emph{Macromolecules} \textbf{2019}, \emph{52},
  7167--7174\relax
\mciteBstWouldAddEndPuncttrue
\mciteSetBstMidEndSepPunct{\mcitedefaultmidpunct}
{\mcitedefaultendpunct}{\mcitedefaultseppunct}\relax
\EndOfBibitem
\bibitem[Sing and Perry(2020)Sing, and Perry]{Sing2020Recent}
Sing,~C.~E.; Perry,~S.~L. Recent progress in the science of complex
  coacervation. \emph{Soft Matter} \textbf{2020}, \emph{16}, 2885--2914\relax
\mciteBstWouldAddEndPuncttrue
\mciteSetBstMidEndSepPunct{\mcitedefaultmidpunct}
{\mcitedefaultendpunct}{\mcitedefaultseppunct}\relax
\EndOfBibitem
\bibitem[Nakamura and Shi(2010)Nakamura, and Shi]{Nakamura2010Self}
Nakamura,~I.; Shi,~A.-C. Self-consistent field theory of polymer-ionic molecule
  complexation. \emph{The Journal of Chemical Physics} \textbf{2010},
  \emph{132}, 194103\relax
\mciteBstWouldAddEndPuncttrue
\mciteSetBstMidEndSepPunct{\mcitedefaultmidpunct}
{\mcitedefaultendpunct}{\mcitedefaultseppunct}\relax
\EndOfBibitem
\bibitem[Nakamura(2016)]{Nakamura2016Spinodal}
Nakamura,~I. Spinodal Decomposition of a Polymer and Ionic Liquid Mixture:
  Effects of Electrostatic Interactions and Hydrogen Bonds on Phase
  Instability. \emph{Macromolecules} \textbf{2016}, \emph{49}, 690--699\relax
\mciteBstWouldAddEndPuncttrue
\mciteSetBstMidEndSepPunct{\mcitedefaultmidpunct}
{\mcitedefaultendpunct}{\mcitedefaultseppunct}\relax
\EndOfBibitem
\bibitem[Sing \latin{et~al.}(2013)Sing, Zwanikken, and de~la
  Cruz]{Sing2013Interfacial}
Sing,~C.~E.; Zwanikken,~J.~W.; de~la Cruz,~M.~O. Interfacial Behavior in
  Polyelectrolyte Blends: Hybrid Liquid-State Integral Equation and
  Self-Consistent Field Theory Study. \emph{Phys. Rev. Lett.} \textbf{2013},
  \emph{111}, 168303\relax
\mciteBstWouldAddEndPuncttrue
\mciteSetBstMidEndSepPunct{\mcitedefaultmidpunct}
{\mcitedefaultendpunct}{\mcitedefaultseppunct}\relax
\EndOfBibitem
\bibitem[Sing \latin{et~al.}(2013)Sing, Zwanikken, and Olvera de~la
  Cruz]{Sing2013Ion}
Sing,~C.~E.; Zwanikken,~J.~W.; Olvera de~la Cruz,~M. Ion Correlation-Induced
  Phase Separation in Polyelectrolyte Blends. \emph{ACS Macro Letters}
  \textbf{2013}, \emph{2}, 1042--1046, PMID: 35581876\relax
\mciteBstWouldAddEndPuncttrue
\mciteSetBstMidEndSepPunct{\mcitedefaultmidpunct}
{\mcitedefaultendpunct}{\mcitedefaultseppunct}\relax
\EndOfBibitem
\bibitem[Pryamitsyn \latin{et~al.}(2017)Pryamitsyn, Kwon, Zwanikken, and Olvera
  de~la Cruz]{Pryamitsyn2017Anomalous}
Pryamitsyn,~V.~A.; Kwon,~H.-K.; Zwanikken,~J.~W.; Olvera de~la Cruz,~M.
  Anomalous Phase Behavior of Ionic Polymer Blends and Ionic Copolymers.
  \emph{Macromolecules} \textbf{2017}, \emph{50}, 5194--5207\relax
\mciteBstWouldAddEndPuncttrue
\mciteSetBstMidEndSepPunct{\mcitedefaultmidpunct}
{\mcitedefaultendpunct}{\mcitedefaultseppunct}\relax
\EndOfBibitem
\bibitem[Sing \latin{et~al.}(2014)Sing, Zwanikken, and Olvera~de
  La~Cruz]{sing2014electrostatic}
Sing,~C.~E.; Zwanikken,~J.~W.; Olvera~de La~Cruz,~M. Electrostatic control of
  block copolymer morphology. \emph{Nature Materials} \textbf{2014}, \emph{13},
  694--698\relax
\mciteBstWouldAddEndPuncttrue
\mciteSetBstMidEndSepPunct{\mcitedefaultmidpunct}
{\mcitedefaultendpunct}{\mcitedefaultseppunct}\relax
\EndOfBibitem
\bibitem[Kong \latin{et~al.}(2021)Kong, Hou, and Qin]{Kong2021Weakening}
Kong,~X.; Hou,~K. J.-Y.; Qin,~J. Weakening of Solvation-Induced Ordering by
  Composition Fluctuation in Salt-Doped Block Polymers. \emph{ACS Macro
  Letters} \textbf{2021}, \emph{10}, 545--550\relax
\mciteBstWouldAddEndPuncttrue
\mciteSetBstMidEndSepPunct{\mcitedefaultmidpunct}
{\mcitedefaultendpunct}{\mcitedefaultseppunct}\relax
\EndOfBibitem
\bibitem[Grzetic \latin{et~al.}(2021)Grzetic, Delaney, and
  Fredrickson]{Grzetic2021Electrostatic}
Grzetic,~D.~J.; Delaney,~K.~T.; Fredrickson,~G.~H. Electrostatic Manipulation
  of Phase Behavior in Immiscible Charged Polymer Blends. \emph{Macromolecules}
  \textbf{2021}, \emph{54}, 2604--2616\relax
\mciteBstWouldAddEndPuncttrue
\mciteSetBstMidEndSepPunct{\mcitedefaultmidpunct}
{\mcitedefaultendpunct}{\mcitedefaultseppunct}\relax
\EndOfBibitem
\bibitem[Fredrickson \latin{et~al.}(0)Fredrickson, Xie, Edmund, Le, Sun,
  Grzetic, Vigil, Delaney, Chabinyc, and Segalman]{Fredrickson2022Ionic}
Fredrickson,~G.~H.; Xie,~S.; Edmund,~J.; Le,~M.~L.; Sun,~D.; Grzetic,~D.~J.;
  Vigil,~D.~L.; Delaney,~K.~T.; Chabinyc,~M.~L.; Segalman,~R.~A. Ionic
  Compatibilization of Polymers. \emph{ACS Polymers Au} \textbf{0}, \emph{0},
  null\relax
\mciteBstWouldAddEndPuncttrue
\mciteSetBstMidEndSepPunct{\mcitedefaultmidpunct}
{\mcitedefaultendpunct}{\mcitedefaultseppunct}\relax
\EndOfBibitem
\bibitem[Wang(2008)]{wang2008Effects}
Wang,~Z.-G. Effects of Ion Solvation on the Miscibility of Binary Polymer
  Blends. \emph{The Journal of Physical Chemistry B} \textbf{2008}, \emph{112},
  16205--16213, PMID: 19007274\relax
\mciteBstWouldAddEndPuncttrue
\mciteSetBstMidEndSepPunct{\mcitedefaultmidpunct}
{\mcitedefaultendpunct}{\mcitedefaultseppunct}\relax
\EndOfBibitem
\bibitem[Nakamura \latin{et~al.}(2011)Nakamura, Balsara, and
  Wang]{Nakamura2011Thermodynamics}
Nakamura,~I.; Balsara,~N.~P.; Wang,~Z.-G. Thermodynamics of Ion-Containing
  Polymer Blends and Block Copolymers. \emph{Phys. Rev. Lett.} \textbf{2011},
  \emph{107}, 198301\relax
\mciteBstWouldAddEndPuncttrue
\mciteSetBstMidEndSepPunct{\mcitedefaultmidpunct}
{\mcitedefaultendpunct}{\mcitedefaultseppunct}\relax
\EndOfBibitem
\bibitem[Zhang \latin{et~al.}(1990)Zhang, Natansohn, and
  Eisenberg]{Zhang1990Intermolecular}
Zhang,~X.; Natansohn,~A.; Eisenberg,~A. Intermolecular cross-polarization
  studies of the miscibility enhancement of PS/PMMA blends through ionic
  interactions. \emph{Macromolecules} \textbf{1990}, \emph{23}, 412--416\relax
\mciteBstWouldAddEndPuncttrue
\mciteSetBstMidEndSepPunct{\mcitedefaultmidpunct}
{\mcitedefaultendpunct}{\mcitedefaultseppunct}\relax
\EndOfBibitem
\bibitem[Zhang and Eisenberg(1990)Zhang, and Eisenberg]{Zhang1990NMR}
Zhang,~X.; Eisenberg,~A. NMR and dynamic mechanical studies of miscibility
  enhancement via ionic interactions in polystyrene/poly (ethyl acrylate)
  blends. \emph{Journal of Polymer Science Part B: Polymer Physics}
  \textbf{1990}, \emph{28}, 1841--1857\relax
\mciteBstWouldAddEndPuncttrue
\mciteSetBstMidEndSepPunct{\mcitedefaultmidpunct}
{\mcitedefaultendpunct}{\mcitedefaultseppunct}\relax
\EndOfBibitem
\bibitem[Hou and Qin(2018)Hou, and Qin]{Hou2018Solvation}
Hou,~K.~J.; Qin,~J. Solvation and Entropic Regimes in Ion-Containing Block
  Copolymers. \emph{Macromolecules} \textbf{2018}, \emph{51}, 7463--7475\relax
\mciteBstWouldAddEndPuncttrue
\mciteSetBstMidEndSepPunct{\mcitedefaultmidpunct}
{\mcitedefaultendpunct}{\mcitedefaultseppunct}\relax
\EndOfBibitem
\bibitem[Hou \latin{et~al.}(2020)Hou, Loo, Balsara, and Qin]{Hou2020Comparing}
Hou,~K.~J.; Loo,~W.~S.; Balsara,~N.~P.; Qin,~J. Comparing Experimental Phase
  Behavior of Ion-Doped Block Copolymers with Theoretical Predictions Based on
  Selective Ion Solvation. \emph{Macromolecules} \textbf{2020}, \emph{53},
  3956--3966\relax
\mciteBstWouldAddEndPuncttrue
\mciteSetBstMidEndSepPunct{\mcitedefaultmidpunct}
{\mcitedefaultendpunct}{\mcitedefaultseppunct}\relax
\EndOfBibitem
\bibitem[Fredrickson(2013)]{FredricksonBook}
Fredrickson,~G. \emph{The Equilibrium Theory of Inhomogeneous Polymers};
  International Series of Monographs on Physics; OUP Oxford, 2013\relax
\mciteBstWouldAddEndPuncttrue
\mciteSetBstMidEndSepPunct{\mcitedefaultmidpunct}
{\mcitedefaultendpunct}{\mcitedefaultseppunct}\relax
\EndOfBibitem
\bibitem[Wang(2010)]{Wang2010Fluctuation}
Wang,~Z.-G. Fluctuation in electrolyte solutions: The self energy. \emph{Phys.
  Rev. E} \textbf{2010}, \emph{81}, 021501\relax
\mciteBstWouldAddEndPuncttrue
\mciteSetBstMidEndSepPunct{\mcitedefaultmidpunct}
{\mcitedefaultendpunct}{\mcitedefaultseppunct}\relax
\EndOfBibitem
\bibitem[de~Gennes(1979)]{de1979scaling}
de~Gennes,~P. \emph{Scaling Concepts in Polymer Physics}; Cornell University
  Press, 1979\relax
\mciteBstWouldAddEndPuncttrue
\mciteSetBstMidEndSepPunct{\mcitedefaultmidpunct}
{\mcitedefaultendpunct}{\mcitedefaultseppunct}\relax
\EndOfBibitem
\bibitem[Nakamura(2014)]{Nakamura2014Ion}
Nakamura,~I. Ion Solvation in Polymer Blends and Block Copolymer Melts: Effects
  of Chain Length and Connectivity on the Reorganization of Dipoles. \emph{The
  Journal of Physical Chemistry B} \textbf{2014}, \emph{118}, 5787--5796, PMID:
  24806716\relax
\mciteBstWouldAddEndPuncttrue
\mciteSetBstMidEndSepPunct{\mcitedefaultmidpunct}
{\mcitedefaultendpunct}{\mcitedefaultseppunct}\relax
\EndOfBibitem
\bibitem[Nakamura \latin{et~al.}(2012)Nakamura, Shi, and Wang]{Nakamura2012Ion}
Nakamura,~I.; Shi,~A.-C.; Wang,~Z.-G. Ion Solvation in Liquid Mixtures: Effects
  of Solvent Reorganization. \emph{Phys. Rev. Lett.} \textbf{2012}, \emph{109},
  257802\relax
\mciteBstWouldAddEndPuncttrue
\mciteSetBstMidEndSepPunct{\mcitedefaultmidpunct}
{\mcitedefaultendpunct}{\mcitedefaultseppunct}\relax
\EndOfBibitem
\bibitem[Nakamura(2015)]{Nakamura2015Dipolar}
Nakamura,~I. Dipolar Self-Consistent Field Theory for Ionic Liquids: Effects of
  Dielectric Inhomogeneity in Ionic Liquids between Charged Plates. \emph{The
  Journal of Physical Chemistry C} \textbf{2015}, \emph{119}, 7086--7094\relax
\mciteBstWouldAddEndPuncttrue
\mciteSetBstMidEndSepPunct{\mcitedefaultmidpunct}
{\mcitedefaultendpunct}{\mcitedefaultseppunct}\relax
\EndOfBibitem
\bibitem[Brown \latin{et~al.}(2018)Brown, Seo, and Hall]{Brown2018Ion}
Brown,~J.~R.; Seo,~Y.; Hall,~L.~M. {Ion correlation effects in salt-doped block
  copolymers}. \emph{Phys. Rev. Lett.} \textbf{2018}, \emph{120},
  127801--1--127801--7\relax
\mciteBstWouldAddEndPuncttrue
\mciteSetBstMidEndSepPunct{\mcitedefaultmidpunct}
{\mcitedefaultendpunct}{\mcitedefaultseppunct}\relax
\EndOfBibitem
\bibitem[Li and Wu(2006)Li, and Wu]{Li2006Density}
Li,~Z.; Wu,~J. Density Functional Theory for Polyelectrolytes near Oppositely
  Charged Surfaces. \emph{Phys. Rev. Lett.} \textbf{2006}, \emph{96},
  048302\relax
\mciteBstWouldAddEndPuncttrue
\mciteSetBstMidEndSepPunct{\mcitedefaultmidpunct}
{\mcitedefaultendpunct}{\mcitedefaultseppunct}\relax
\EndOfBibitem
\bibitem[Wu and Li(2007)Wu, and Li]{Wu2007Density}
Wu,~J.; Li,~Z. Density-Functional Theory for Complex Fluids. \emph{Annual
  Review of Physical Chemistry} \textbf{2007}, \emph{58}, 85--112, PMID:
  17052165\relax
\mciteBstWouldAddEndPuncttrue
\mciteSetBstMidEndSepPunct{\mcitedefaultmidpunct}
{\mcitedefaultendpunct}{\mcitedefaultseppunct}\relax
\EndOfBibitem
\bibitem[Kong \latin{et~al.}(2017)Kong, Lu, Wu, and Liu]{Kong2017SLB}
Kong,~X.; Lu,~D.; Wu,~J.; Liu,~Z. A theoretical study on the morphological
  phase diagram of supported lipid bilayers. \emph{Phys. Chem. Chem. Phys.}
  \textbf{2017}, \emph{19}, 16897--16903\relax
\mciteBstWouldAddEndPuncttrue
\mciteSetBstMidEndSepPunct{\mcitedefaultmidpunct}
{\mcitedefaultendpunct}{\mcitedefaultseppunct}\relax
\EndOfBibitem
\end{mcitethebibliography}

\end{document}